\documentclass[
 reprint,
nofootinbib,
 amsmath,amssymb,
 aps,superscriptaddress
]{revtex4-2}

\usepackage{graphicx, color}
\usepackage{dcolumn}
\usepackage{bm}
\usepackage[mathlines]{lineno}
\usepackage{amsmath}
\usepackage{amssymb}
\usepackage{bm}
\usepackage{mathrsfs}
\usepackage{float}
\usepackage{soul}
\usepackage{mathrsfs}
\usepackage{lipsum}
\usepackage{hyperref}
\usepackage[normalem]{ulem}
\hypersetup{
    colorlinks=true,       
    linkcolor=green2,          
    citecolor=green2,        
    filecolor=magenta,      
    urlcolor=green2           
}
\usepackage{cleveref}
\usepackage{color}

\newcommand{\mpl}{m_\mathrm{pl}}

\definecolor{rp}{cmyk}{0.2, 1, 0.6, 0}
\definecolor{rp}{cmyk}{0.2, 1, 0.6, 0}
\definecolor{green2}{cmyk}{0.27, 0, 1, 0.52}

\newcommand{\schr}{\rm Schr{\"o}dinger }

\def\mpl{m_\mathrm{pl}}

\usepackage{hyperref}

\begin{document}

\preprint{APS/123-QED}

\title{Kinetic relaxation and nucleation of Bose stars in self-interacting wave dark matter}

\author{Mudit Jain}
\email{mudit.jain@rice.edu}
\affiliation{Department of Physics and Astronomy, Rice University, Houston, Texas 77005, U.S.A.}
\affiliation{Theoretical Particle Physics and Cosmology, King's College London, Strand, London, WC2R 2LS, United Kingdom}
\author{Wisha Wanichwecharungruang}
\email{wisha@rice.edu}
\author{Jonathan Thomas}
\email{jonathan.thomas@rice.edu}

\affiliation{Department of Physics and Astronomy, Rice University, Houston, Texas 77005, U.S.A.}

\date{\today}

\begin{abstract}
We revisit kinetic relaxation and soliton/Boson star nucleation in fuzzy scalar dark matter featuring short-ranged self-interactions $\mathcal{H}_{\rm int} = -\lambda|\psi|^4/2m^2$, alongside gravitational self-interactions. We map out the full curve of nucleation timescale for both repulsive ($\lambda < 0$) and attractive ($\lambda > 0$) short-ranged self-interaction strength, and in doing so reveal two new points. Firstly, besides the two usual terms, $\propto G^2$ and $\propto \lambda^2$, in the total relaxation rate $\Gamma_{\rm relax}$, there is an additional cross term $\propto G\lambda$ arising due to interference between gravitational and short-ranged self-interaction scattering amplitudes. This yields a critical repulsive interaction strength $\lambda_{\rm cr} \simeq - 2\pi Gm^2/v_{0}^2$, at which the relaxation rate is smallest and serves as the transition point between 
typical net attractive self-interaction ($\lambda \gtrsim \lambda_{\rm cr}$), and net repulsive self-interaction ($-\lambda \gtrsim -\lambda_{\rm cr}$). Secondly, while in the net attractive regime, nucleation time scale is similar to inverse relaxation time scale $\tau_{\rm nuc} \sim \Gamma^{-1}_{\rm relax}$, 
in the net repulsive regime nucleation occurs at a delayed time $\tau_{\rm nuc} \sim (\lambda/\lambda_{\rm cr})\Gamma^{-1}_{\rm relax}$. We confirm our analytical understanding by performing 3D field simulations with varying average mass density $\bar{\rho}$, box size $L$ and grid size $N$. 
\end{abstract}

\maketitle

\tableofcontents

\section{Introduction}\label{sec:intro}

Understanding the nature of dark matter (DM) is one of the main quests of modern cosmology. It could be multi-faceted in the sense that there are many degrees of freedom in the whole dark sector, for instance the String theory Axiverse~\cite{Arvanitaki:2009fg,Acharya:2010zx,Cicoli:2012sz} or other confined sector(s) (e.g. see~\cite{Giedt:2000bi,Taylor:2015ppa}, and also~\cite{Alexander:2023wgk}). Or it could be that there is a dominant degree of freedom, such as the QCD axion~\cite{Peccei:1977hh,Weinberg:1977ma,Wilczek:1977pj,Preskill:1982cy,Abbott:1982af,Dine:1982ah}, that comprises all (or most) of the dark matter. Furthermore, while the DM appears to interact only gravitationally with the Standard Model degrees of freedom (or very weakly if it does otherwise), it can still have appreciable non-gravitational self-interactions (nGSI) besides the usual gravitational self-interactions (GSI). Such is the case even for the above mentioned examples.\\



For bosonic particles (of any integer spin) and high enough occupation numbers, which is indeed the case for particle masses below a few eV, classical description of the associated field suffices and the dynamics is described by a non-linear \schr equation in the non-relativistic regime. The non-linear \schr equation entails novel wave dynamics owing to the de-Broglie scale becoming manifestly important. As a few examples, suppression of structure on small scales~\cite{Matos:2000ss,Hu:2000ke}, turbulence~\cite{Mocz:2017wlg}, superradiance~\cite{Brito:2015oca}, vortices~\cite{Schive:2014dra}, bound states called solitons/Bose stars~\cite{Ruffini:1969qy,Kolb:1993hw,Chavanis:2011zi,Amin:2011hj,Schive:2014dra,Brito:2015pxa,Aoki:2017ixz,Amin:2019ums,Adshead:2021kvl,Jain:2021pnk,Gorghetto:2022sue,Amin:2022pzv}, interference patterns~\cite{Hui:2020hbq,Amin:2022pzv,Gosenca:2023yjc}, field correlation scales depending upon the nature of self-interaction~\cite{Guth:2014hsa}, etc. For comprehensive recent reviews in the case of scalar DM, see~\cite{Ferreira:2020fam,Hui:2021tkt}.

Of particular interest to us in this paper, is the phenomenon of kinetic relaxation and associated nucleation of Bose stars within a bath of DM waves~\cite{Semikoz:1994zp,Levkov:2018kau,Eggemeier:2019jsu,Kirkpatrick:2020fwd,Chen:2021oot,Kirkpatrick:2021wwz,Purohit:2021hpg,Jain:2023ojg,Chen:2023bqy}. The term ``kinetic" implies two key aspects: (a) The self-interactions in the field are small, allowing wave modes to freely evolve (at leading order) with the non-relativistic dispersion relation $\omega_k = k^2/2m$. This enables a kinetic treatment of the mode occupation number function; (b) the size of the `box' ($\sim$ the size of a DM halo for practical purposes), is much larger than the typical fluctuation scale $\ell_{\rm dB} \sim \pi/\bar{k} \sim \pi/(mv_{0})$ in the bath of DM waves. The process of kinetic relaxation is attributed to these self-interactions of the DM field which although small, over large time scales $\tau_{\rm relax} \gg \omega_{\bar k}^{-1}$ drive the occupation number function to develop increasing support towards smaller wavenumbers ${\bm k} \rightarrow 0$. See~\cite{Semikoz:1994zp} and~\cite{Jain:2023ojg} for a relevant discussion for the cases of point-like quartic self-interactions and gravitational self-interactions respectively. Once enough particles condense into lower momentum states, their collective \textit{net attractive} self-interaction becomes strong enough to counter balance their wave pressure resulting in the nucleation of a Bose star.\\

In this paper, we focus on investigating kinetic relaxation and subsequent Bose star nucleation for a single scalar \schr field with both GSI and point-like quartic nGSI. Employing wave-kinetic Boltzmann analysis and 3D simulations, we demonstrate the presence of a previously overlooked cross term $\propto G\lambda$ in the rate of \textit{relaxation} $\Gamma_{\rm relax}$. (Here $G$ denotes Newton's constant and $\lambda$ represents the point-like nGSI strength). It arises due to 
interference between the gravitational and point-like self-interaction scattering amplitudes. The presence of this cross term gives rise to a critical nGSI (repulsive) strength $\lambda_{\rm cr} \simeq -(2\pi G)m^2/v_0^2$, at which the rate of relaxation reaches its minimum value (corresponding to maximum nucleation time). This critical value also serves as the transition point from typical net (contributions from both gravity and short-ranged self-interactions) attractive to repulsive self-interactions.

Because of the presence of gravitational self-interaction, kinetic relaxation is generally accompanied with \textit{nucleation} of spatially localized clumps/Bose stars, with their nucleation times dependent on the nature of the short-ranged self-interactions -- attractive or repulsive. For $\lambda \gtrsim \lambda_{\rm cr}$, the net typical self-interaction is attractive, and nucleation happens quickly after relaxation. On the other hand for $\lambda \lesssim \lambda_{\rm cr}$, the net typical self-interaction is repulsive and nucleation gets delayed. We will study relaxation and nucleation of Bose stars, and also discuss their eventual fate.\footnote{Following conventional nomenclature, we shall use the words relaxation and condensation interchangeably, but it is to be stressed that nucleation (of a bound state) is not always equivalent to relaxation/condensation. As we shall see, it is equivalent to the other two in the net attractive regime, whereas different in the net repulsive regime.} However we will not dwell into a careful analysis of the growth rate of these nucleated stars. See~\cite{Chen:2020cef,Chan:2022bkz,Dmitriev:2023ipv} for the gravity only ($\lambda = 0$) case.\\

The rest of the paper is organized as follows: Starting with the basic model of fuzzy scalar DM carrying both GSI and point-like nGSI in sec.~\ref{sec:model}, we describe the associated wave kinetic Boltzmann equation for the evolution of the occupation number function in sec.~\ref{sec:Boltzmann_analysis}. Highlighting the presence of the cross term (that gives rise to $\lambda_{\rm cr}$), we estimate the total rate of kinetic relaxation/condensation. In sec.~\ref{sec:nnucleation_behavior} we discuss the two cases of $\lambda \gtrsim \lambda_{\rm cr}$ and $-\lambda \gtrsim -\lambda_{\rm cr}$ and write down the associated nucleation time scales of spatially localized bound objects. In sec.~\ref{sec:simulations} we discuss our 3D simulations and compare our analytical estimates with them. We also discuss eventual behavior of Bose clumps observed in simulations. Finally in sec.\ref{sec:summary}, we summarize our work and also compare our results with the existing literature on this subject. In appendix~\ref{sec:convergence} we discuss statistical convergence of our simulations, and in appendix~\ref{sec:peculiarity} we discuss a peculiarity observed in the case of repulsive short-ranged self-interactions, over longer time scales as compared to nucleation.\\

\noindent{{\bf Conventions}}: Unless stated otherwise, we will work in units where $\hbar = c = 1$.

\section{Model}\label{sec:model}

Ignoring Hubble flow (for we are interested in sufficiently sub-horizon dynamics), the evolution of the cold/non-relativistic fuzzy scalar dark matter with both GSI and short-ranged quartic nGSI, can be described using mean field theory. The dark matter field $\psi$ obeys the following non-linear \schr (Gross-Pitaevskii) equation:
\begin{align}\label{eq:SP_physical}
    &i\frac{\partial}{\partial t}\psi = -\frac{1}{2m}\nabla^2\psi + \psi\Bigl(4\pi G m^2\nabla^{-2}_{/0} - \frac{\lambda}{m^2}\Bigr)\psi^{\ast}\psi\,.
\end{align}
Here $G$ is the Newton's constant, and $\lambda$ is the point like self-interaction strength. In our convention, $\lambda > 0$ and $\lambda < 0$ dictate attractive and repulsive self-interaction respectively. To obtain the form Eq.~\eqref{eq:SP_physical}, we have plugged the self-gravitational potential, $\Phi = 4\pi G\,\nabla^{-2}(m\psi^{\ast}\psi - \bar{\rho}) \equiv 4\pi Gm\nabla^{-2}_{/0}\psi^{\ast}\psi$, in the usual Schr\"{o}dinger-Poisson system of equations. The $\nabla^{-2}_{/0}$ denotes exclusion of the homogeneous part of the number density field $\psi^{\ast}\psi$. In Fourier space with the decomposition $\psi({\bm x},t) = (2\pi)^{-3}\int\mathrm{d}{\bm k}\,e^{-i{\bm k}\cdot{\bm x}}\,\Psi_{\bm k}(t)$, the \schr equation becomes
\begin{align}\label{eq:SP_Fourier}
    i\dot{\Psi}_{\bm k} &= \frac{k^2}{2m}\Psi_{\bm k} + \int\frac{\mathrm{d}{\bm p}}{(2\pi)^3}\frac{\mathrm{d}{\bm q}}{(2\pi)^3}\frac{\mathrm{d}{\bm \ell}}{(2\pi)^3}\mathcal{T}_{\bm k, \bm p, \bm q, \bm \ell}\,\Psi^{\ast}_{\bm p}\Psi_{\bm q}\Psi_{\bm \ell}\nonumber\\
    &\qquad\qquad\qquad\qquad\times\,(2\pi)^3\delta^{(3)}(\bm k + \bm p - \bm q - \bm \ell)\,,
\end{align}
where
\begin{align}\label{eq:formfactor}
    \mathcal{T}_{\bm k, \bm p, \bm q, \bm \ell} =& -\frac{4\pi G m^2}{|{\bm k}-{\bm \ell}|^{2}} - \frac{\lambda}{m^2}\,,
\end{align}
and it is understood that ${\bm k} \neq {\bm \ell} \neq 0$ in the above. For later convenience, it is also useful to write down the Hamiltonian density (in physical space) for the mean field $\psi$:
\begin{align}\label{eq:Hamiltonian_density}
    \mathcal{H} = \frac{1}{2m}|\nabla\psi|^2 + m\Phi|\psi|^2 - \frac{\lambda}{2m^2}|\psi|^4\,.
\end{align}
Here the different terms in the above can be attributed to the wave-pressure $\mathcal{H}_{\rm wp} = |\nabla\psi|^2/2m$, gravitational self-interaction $\mathcal{H}_{\rm gr} = m\Phi|\psi|^2$, and short-ranged self-interaction $\mathcal{H}_{\rm self} = -\lambda|\psi|^4/2m^2$.\\

The Gross-Pitaevskii (GP) equation, being non-linear, renders it difficult to analyze and study the evolution of the $\psi$ field in generality. However for the purposes of kinetic relaxation leading to nucleation of localized Bose clumps however, wave kinetic Boltzmann analysis can be performed which we discuss next. To test and verify our analytical understanding, we perform 3D field simulations which we discuss in a later section. 

\section{Wave kinetics and relaxation}\label{sec:Boltzmann_analysis}

For kinetic relaxation in wave dynamics, we can study the evolution of the mode occupation number function $f_{\bm k} = |\Psi_{\bm k}|^2/V$ ($V$ is the volume), which is nothing but the Fourier transform of the $2$-point volume averaged field correlator $\zeta({\bm x},t) = V^{-1}\int\mathrm{d}{\bm y}\,\psi^{\ast}(\bm{y},t)\,\psi(\bm{y} + \bm{x},t)$. Under random phase approximation with weak interactions, the relevant wave-kinetic Boltzmann equation can be derived. See for instance~\cite{Zakharov1992KolmogorovSO}. For a derivation for the general case of arbitrary number of fields and $2$ body interactions, see~\cite{Jain:2023ojg}. For the scalar case at hand, characterizing the dependence of the occupation number functions on wavenumbers as $f_{\bm k/m}$, the wave-kinetic equation takes the familiar form
\begin{widetext}
    \begin{align}\label{eq:Boltzmann_general}
    &\frac{\partial f_{\bm k/m}}{\partial t} = \int\frac{\mathrm{d}{\bm p}}{(2\pi)^3}\,\mathrm{d}\sigma_{{\bm k} + {\bm p} \rightarrow {\bm q} + {\bm \ell}}\,|{\bm v} - \tilde{\bm v}|\,\Biggl[(f_{\bm k/m} + f_{\bm p/m})f_{\bm q/m}f_{\bm \ell/m} - (f_{\bm q/m} + f_{\bm \ell/m})f_{\bm k/m}f_{\bm p/m}\Biggr]\,,
    \nonumber\\
    &{\rm where} \quad \mathrm{d}\sigma_{{\bm k} + {\bm p} \rightarrow {\bm q} + {\bm \ell}} = \frac{1}{2|{\bm v} - \tilde{\bm v}|}\frac{\mathrm{d}{\bm q}}{(2\pi)^3}\frac{\mathrm{d}{\bm \ell}}{(2\pi)^3}\Bigl(\mathcal{T}_{\bm k, \bm p, \bm q, \bm \ell} + \mathcal{T}_{\bm k, \bm p, \bm \ell, \bm q}\Bigr)\Bigl(\mathcal{T}_{\bm k, \bm p, \bm q, \bm \ell} + \mathcal{T}_{\bm k, \bm p, \bm \ell, \bm q}\Bigr)^{\ast}\,\times\nonumber\\
    &\qquad\qquad\qquad\qquad\qquad\qquad\qquad\qquad\qquad\qquad\qquad\qquad\qquad (2\pi)^4\delta^{(3)}(\bm k + \bm p - \bm q - \bm \ell)\,\delta(E_{\bm k} + E_{\bm p} - E_{\bm q} - E_{\bm \ell})\,.
    \end{align}
\end{widetext}

Here ${\bm v} = {\bm k}/m$ and $\tilde{\bm v} = {\bm p}/m$ are the incoming ``velocities" in the $2$-wave interaction, and the quantities in the $1$-dimensional Dirac delta function are the free wave energies $E_{\bm k} = k^2/2m$. The quantity $\mathrm{d}\sigma$ is the effective differential cross section. The cubic nature of the terms in the right hand side bracket ($\sim f_{\bm x}f_{\bm y}f_{\bm z}$), usually understood as Boltzmann enhancement terms, arise due to the wave-mechanical nature of the system~\eqref{eq:SP_physical} and are crucial for the phenomenon of Bose condensation. Last but not the least, it is the form of the differential cross section that appears in the wave-kinetic equation, $\sim |\mathcal{T}|^2$, that is of utmost importance for our discussion. The scattering amplitudes due to the different kinds of $2$-body interactions (here gravity and point-like self-interactions), are \textit{added first and then squared}: What appears in the differential cross section is $|\mathcal{T}|^2$ where $\mathcal{T} = \mathcal{T}_{G} + \mathcal{T}_{\lambda}$ (c.f. Eq.~\eqref{eq:formfactor}), and $|\mathcal{T}_G + \mathcal{T}_{\lambda}|^2 \neq |\mathcal{T}_G|^2 + |\mathcal{T}_{\lambda}|^2$ (since both $\mathcal{T}_G \propto -4\pi G m^2$ and $\mathcal{T}_{\lambda} = -\lambda/m^{2}$ are real). This can be attributed to the wave dynamical nature of the GP system. The above equation~\eqref{eq:Boltzmann_general}, after integration over the Dirac deltas, can be re-written in terms of the incoming and outgoing relative velocities ${\bm u} = u\hat{\bm n}$ and ${\bm u}' = u\hat{\bm n}'$ respectively\footnote{Note that the magnitude of the relative velocity does not change in an elastic collision, i.e. $|{\bm u}| = |{\bm u}'| \equiv u$.}, by redefining ${\bm p}/m = {\bm k}/m - u\hat{{\bm n}}$ and ${\bm q}/m = {\bm \ell}/m - u\hat{\bm n}'$:
\begin{widetext}
\begin{align}\label{eq:Boltzmann_grav+SI}
    \frac{\partial f_{\bm v}}{\partial t} &= m^3\int\frac{\mathrm{d}{\bm u}}{(2\pi)^3}\,\mathrm{d}\sigma\,u\Biggl[(f_{\bm v} + f_{\tilde{\bm v}})f_{\tilde{\bm v} - {\bm w}}f_{\bm v + \bm w} - (f_{\tilde{\bm v} - \bm w} + f_{\bm v + \bm w})f_{\bm v}f_{\tilde{\bm v}}\Biggr]\,,\nonumber\\
    {\rm where} \quad \mathrm{d}\sigma &= \frac{\mathrm{d}\Omega_{n'}}{32\pi^2m^2}\Biggl[\Bigl(\frac{16\pi m^2G}{u^2|\hat{\bm n}' - \hat{\bm n}|^2} + \lambda\Bigr)^2 + \Bigl(\frac{16\pi m^2G}{u^2|\hat{\bm n}' + \hat{\bm n}|^2} + \lambda\Bigr)^2+ 2\Bigl(\frac{16\pi m^2G}{u^2|\hat{\bm n}' - \hat{\bm n}|^2} + \lambda\Bigr)\Bigl(\frac{16\pi m^2G}{u^2|\hat{\bm n}' + \hat{\bm n}|^2} + \lambda\Bigr)\Biggr]\,.
\end{align}
\end{widetext}

\begin{figure}[!t]
    \centering
\includegraphics[width=0.48\textwidth]{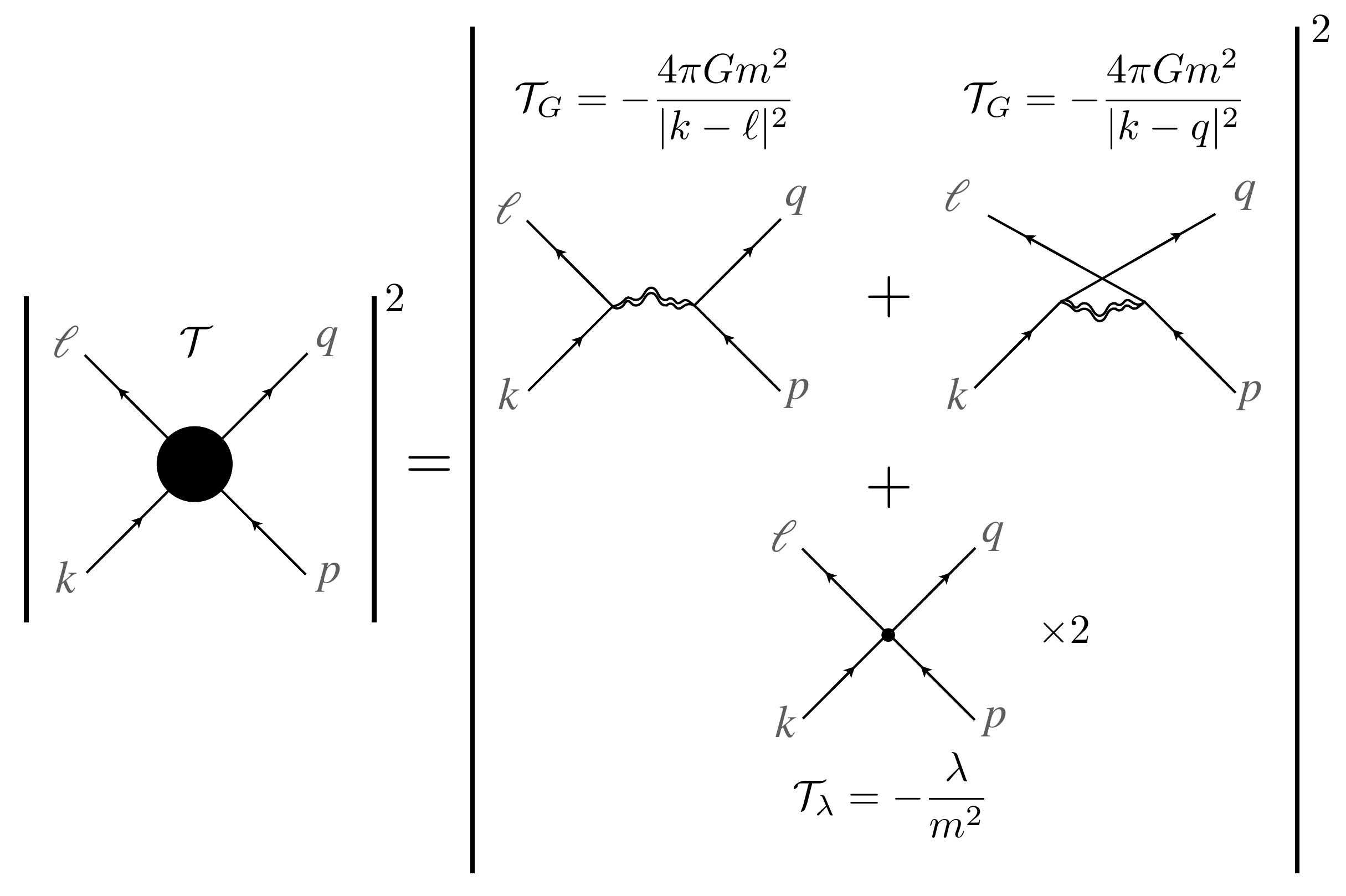}
    \caption{Pictorial / Feynman graph representation of all the terms appearing in the differential cross section in Eq.~\eqref{eq:Boltzmann_grav+SI}. The total contribution to the interaction rate (left hand side of the equality), is the square of the sum of both gravitational amplitude (top two graphs) and point self-interaction amplitude (bottom graph $\times 2$). For gravitational interaction, there are two distinct channels ($t$ and $u$). Their mutual interference, as compared to their sole contributions, becomes subdominant in the large log limit (leading to Rutherford scattering result). More importantly, interference between scattering amplitudes of the two different interactions matters. See main text for details.} 
    \label{fig:graphs}
\end{figure}

Let us briefly discuss the different terms in the differential cross section explicitly. Broadly speaking, there are two types of interference terms that arise. One is the interference between the $t$ and $u$ channels (relevant mainly for the gravitational interaction), and the second is the interference between the two different types of interactions (gravitational and short-ranged). See fig.~\ref{fig:graphs} for a pictorial representation.

For GSI only ($\lambda = 0$) case, the contribution from the $t$ and $u$ channels are the first two terms $\propto G^2\,|\hat{\bm n}' - \hat{\bm n}|^{-4}$ and $\propto G^2\,|\hat{\bm n}' + \hat{\bm n}|^{-4}$, whereas the second term $\propto G^2\,|\hat{\bm n}' - \hat{\bm n}|^{-2}\,|\hat{\bm n}' + \hat{\bm n}|^{-2}$ is due to their mutual interference (as also discussed in~\cite{Jain:2023ojg}). Note that the sole contributions from the $t$ and $u$ channels are identical: The full integral with the $|\hat{\bm n}' + \hat{\bm n}|^{-4}$ term is identical to that with the $|\hat{\bm n}' - \hat{\bm n}|^{-4}$ term. The sole contributions give rise to the Rutherford scattering cross section, carrying a logarithmic IR divergence (aka the Coulomb logarithm), while the interference term becomes sub-dominant in the large log limit and can be omitted. 

For nGSI only ($G = 0$) case, contributions from $t$ and $u$ channels are identical to their mutual interference one, and goes as $\lambda^2$. This is simply due to the interaction being a contact/point interaction.

Importantly when both of the interactions are present, their respective scattering amplitudes (for either of the two channels) are added first and then squared. All the terms $\propto G\lambda$, while giving identical contributions, characterize the interference between the two types of interactions. Splitting the contributions from GSI, nGSI, and their interference, we have the following wave-kinetic Boltzmann equation
\begin{widetext}
\begin{align}\label{eq:Boltzmann_grav+SI_simplified}
    \frac{\partial f_{\bm v}}{\partial t} &= \mathcal{C}_{\rm GSI} + \mathcal{C}_{\rm cross} + \mathcal{C}_{\rm nGSI} \nonumber\\
    \qquad \qquad {\rm where} \quad \mathcal{C}_{\rm GSI} &=  \frac{\Lambda(4\pi G)^2m^5}{4\pi}\nabla_{{v}^i}\Biggl[\frac{1}{2}\nabla_{v^j}f_{{\bm v}}\int\frac{\mathrm{d}\tilde{\bm v}}{(2\pi)^3}\,f_{\tilde{\bm v}}\,\frac{\delta_{ij} - \hat{u}_i\hat{u}_j}{u}\,f_{\tilde{\bm v}} + f_{{\bm v}}\,f_{{\bm v}}\int\frac{\mathrm{d}\tilde{\bm v}}{(2\pi)^3}\,\frac{\hat{u}_i}{u^2}\,f_{\tilde{\bm v}}\Biggr],\nonumber\\
    \mathcal{C}_{\rm cross} &= \frac{(4\pi G)\lambda m^3}{4\pi}\int\frac{\mathrm{d}\Omega_n}{4\pi}\frac{\mathrm{d}u\,u^2}{2\pi^2}\frac{\mathrm{d}\Omega_{n'}}{4\pi}\frac{u}{|{\bm w}|^2}\Biggl[(f_{\bm v} + f_{\tilde{\bm v}})f_{\bm v + {\bm w}}f_{\tilde{{\bm v}} - {\bm w}} - (f_{\bm v + {\bm w}} + f_{\tilde{\bm v} - {\bm w}})f_{\bm v}f_{\tilde{\bm v}}\Biggr],\nonumber\\
    \mathcal{C}_{\rm nGSI} &= \frac{\lambda^2 m}{2\pi}\int\frac{\mathrm{d}\Omega_n}{4\pi}\frac{\mathrm{d}u\,u^2}{2\pi^2}\frac{\mathrm{d}\Omega_{n'}}{4\pi}\,u\Biggl[(f_{\bm v} + f_{\tilde{\bm v}})f_{\bm v + {\bm w}}f_{\tilde{{\bm v}} - {\bm w}} - (f_{\bm v + {\bm w}} + f_{\tilde{\bm v} - {\bm w}})f_{\bm v}f_{\tilde{\bm v}}\Biggr]\,.
\end{align}
\end{widetext}
Here ${\bm w} = u(\hat{\bm n}'-\hat{\bm n})/2$, and $\Lambda = \log(m v_0 L)$ is the aforementioned Coulomb logarithm with $v_0$ and $L$ equal to typical velocity and box size (or halo size for physical considerations) respectively. While the cross term and nGSI term follow straightforwardly from Eq.~\eqref{eq:Boltzmann_grav+SI}, the Rutherford scattering collision term $\mathcal{C}_{\rm GSI}$ is obtained after an eikonal approximation and was derived explicitly in~\cite{Jain:2023ojg}. Also see~\cite{Zakharov1992KolmogorovSO,Chavanis:2020upb} for the same equation for a scalar field.

Now in order to get a typical estimate for the total relaxation rate $\Gamma_{\rm relax} \equiv \frac{1}{f_{\bm v}}\frac{\partial f_{\bm v}}{\partial t}$, we can replace different quantities in the three collision terms with their appropriate scalings. Replacing angular volume $\int\mathrm{d}\Omega \rightarrow 4\pi$, typical relative velocity $|\hat{\bm n}'-\hat{\bm n}| \rightarrow \sqrt{2}$, velocity derivative $\nabla_{v} \rightarrow 1/v_0$, velocity integral $\int\mathrm{d}u\,u^{n-1} \rightarrow v_0^{n}/n$, and finally the occupation number function $f_{\bm v} \rightarrow (2\pi)^{3/2}\bar{\rho}/(m^4v_0^3)$, the total relaxation rate is parameterized as
\begin{align}\label{eq:rate_master}
    \Gamma_{\rm relax} \simeq \alpha_1\frac{(4\pi G)^2\bar{\rho}^2\Lambda}{4m^3v_0^6} + \alpha_{12}\frac{(4\pi G)\lambda\bar{\rho}^2}{m^5v_0^4} + \alpha_{2}\frac{\lambda^2\bar{\rho}^2}{m^7v_0^2}\,.
\end{align}
Our scaling of the occupation number is dictated by Gaussian initial condition (see Eq.~\eqref{eq:Gaussian_initialansatz} ahead) which we shall use to perform simulations, described in the next section. In general, $\alpha_{1}$, $\alpha_{12}$, and $\alpha_{2}$ are positive $\mathcal{O}(1)$ coefficients that would depend on the specific initial conditions. Eq.~\eqref{eq:rate_master} is our master formula for the relaxation rate. The value of $\lambda$ around which the relaxation rate becomes smallest, is easily estimated to be
\begin{align}\label{eq:lambda_cr}
    \lambda_{\rm cr} = -\beta\frac{2\pi G\,m^2}{v_0^2} \sim 10^{-57}\left(\frac{10^{-4}}{v_0}\right)^2\left(\frac{m}{10^{-5}\,{\rm eV}}\right)^2\,,
\end{align}
where $\beta = \alpha_{12}/\alpha_2 \sim \mathcal{O}(1)$, and the associated (minimum) rate is\footnote{In the large Coulomb limit (relevant for realistic scenarios), $\Lambda = \log(mv_0L) \gg 1$, and the rate is always positive.}
\begin{align}
    \Gamma_{\rm relax}\left(\lambda_{\rm cr}\right) \simeq \frac{(4\pi G)^2\bar{\rho}^2}{4m^3v_0^6}\left(\alpha_1\Lambda - \frac{\alpha_{12}^2}{\alpha_2}\right)\nonumber\,.
\end{align}
Notice that this critical value of $\lambda_{\rm cr}$, can also be obtained from the GP equation~\eqref{eq:SP_physical} by balancing the gravitational term with the self-interaction term together with replacing the exchange momenta by its typical value $|{\bm k}-{\bm \ell}|^2 \sim 2(mv_0)^2$. This criticality marks the transition point from attractive to repulsive net typical self-interactions: For $\lambda \gtrsim \lambda_{\rm cr}$, typical interactions within the bath of DM waves are attractive since typical $\mathcal{T}$ is negative, whereas for $-\lambda \gtrsim -\lambda_{\rm cr}$ they are repulsive since typical $\mathcal{T}$ is positive.\\ 


\section{Nucleation and behavior of solitons}\label{sec:nnucleation_behavior}

\subsection{Nucleation}

In general, the process of kinetic relaxation is characterized by an increasing support of the occupation number function $f_{\bm k}$ at vanishing wavenumber. (For instance see~\cite{Semikoz:1994zp} and~\cite{Jain:2023ojg} for discussions of nGSI and GSI cases respectively). This implies increasing field correlation over larger length scales with diminishing density fluctuations, i.e. field homogenization. A heuristic understanding of the subsequent nucleation of a spatially localized and bound clump, can perhaps be gained most easily from a particle physics perspective, together with recalling that $\lambda_{\rm cr}$ also marks the transition from typical net attractive self-interaction to typical net repulsive self-interaction. As particles lose kinetic energy on account of self-interactions and move towards smaller momenta (condensate state), there comes a time when within some region, the collective net potential (due to both self-gravitational and short-ranged interactions) becomes comparable to wave pressure. The time scale of this process is nothing but the inverse relaxation rate Eq.~\eqref{eq:rate_master}, which in the case of net attractive self-potential $\lambda \gtrsim \lambda_{\rm cr}$, leads to `immediate locking' of such a region into a bound clump (having negative energy). That is, $ \tau_{\rm nuc} \simeq \Gamma_{\rm relax}^{-1}$. Strictly speaking, this can be taken as a definition of $\tau_{\rm nuc}$ with $\Gamma_{\rm nuc} = \Gamma_{\rm relax}$, in which case the different $\alpha$ constant coefficients in the rate Eq.~\eqref{eq:rate_master}, are understood as such.

On the other hand for $-\lambda \gtrsim -\lambda_{\rm cr}$, relaxation cannot immediately lead to nucleation of a localized bound clump. This is because the net typical interaction is repulsive: the collective self-potential within density fluctuation regions is not binding yet. Over time though, more particles get driven towards the condensate phase, and eventually there arises a potential for a bound object to nucleate (within which net gravity can now compensate for both repulsive short-ranged interaction and wave-pressure). This gives $\tau_{\rm nuc} > \Gamma_{\rm relax}^{-1}$. In general, we can therefore write the following
\begin{align}\label{eq:taunuc}
    \tau_{\rm nuc} \simeq \frac{1}{\Gamma_{\rm relax}}
    \begin{cases}
        1 \qquad & \lambda \gtrsim \lambda_{\rm cr}\\
        h(\lambda) \qquad & -\lambda \gtrsim -\lambda_{\rm cr}
    \end{cases}
\end{align}
where $h(\lambda)$ is a threshold function (or the delay factor), that relates nucleation times to relaxation rates.  As mentioned earlier, relaxation means field homogenization, and we expect the rate at which the system relaxes to be comparable to the rate at which density fluctuations decrease.
The delay factor can then be estimated as the ratio of typical density fluctuation at relaxation, to that at nucleation, $h \sim \delta\rho_{\rm relax}/\delta\rho_{\rm nuc}$. While we expect this to be order unity for net attractive case (first case of Eq.~\ref{eq:taunuc}), for large repulsive strengths it should increase with increasing $-\lambda$. Below we estimate this scaling.

Consider a region of typical size $\sim (mv_0)^{-1}$ where the field would have `locked' itself into a bound configuration upon relaxation/condensation, were the net potential was binding. However this is not the case yet, and we may balance the typical self-interaction energy density (mostly due to short-ranged interactions) $\mathcal{H}_{\rm self} \sim -\lambda\,\delta\rho_{\rm relax}^2/2m^4$, with the wave pressure within $\mathcal{H}_{\rm wp} \sim v_0^2\delta\rho_{\rm relax}/2$. This gives $\delta\rho_{\rm relax} \sim m^4v_0^2/\lambda$. As relaxation continues (meaning more particles are driven towards low momenta state), the value of both density fluctuations $\delta\rho$ and typical size of fluctuation regions $(mv)^{-1}$ change. The former decreases and the latter increases so as to maintain $\mathcal{H}_{\rm self} \sim \mathcal{H}_{\rm wp}$. Then, nucleation is expected to occur when gravity can compensate for both the wave pressure and repulsive short-ranged self-interaction. That is, we may balance (the magnitudes of) all the three energy densities, $\mathcal{H}_{\rm wp} \sim v^2_{\rm nuc}\delta\rho_{\rm nuc}/2$, $|\mathcal{H}_{\rm gr}| \sim 2\pi G\delta\rho^2_{\rm nuc}/(m v_{\rm nuc})^2$, and $\mathcal{H}_{\rm self} \sim -\lambda\delta\rho^2_{\rm nuc}/2m^4$, to give $\delta\rho_{\rm nuc} \sim m^2v^4_{\rm nuc}/(4\pi G)$ and $v_{\rm nuc} \sim (4\pi G m^2/(-\lambda))^{1/2}$. Eliminating $v_{\rm nuc}$ from $\delta\rho_{\rm nuc}$ gives the following estimate for the delay factor
\begin{align}\label{eq:hfunc}
    h(\lambda) \sim \frac{\delta\rho_{\rm relax}}{\delta\rho_{\rm nuc}} \rightarrow \alpha_3\left(\frac{\lambda}{\lambda_{\rm cr}}\right)\,.
\end{align}
Here we have inserted another constant coefficient $\alpha_3$ that depends upon the initial conditions. Through simulations, we will confirm our estimate Eq.~\eqref{eq:taunuc} (together with Eq.~\eqref{eq:rate_master} and Eq.~\eqref{eq:hfunc}), and also extract the different $\alpha$ coefficients for Gaussian initial conditions.

\subsection{Eventual behavior}\label{sec:eventualbehavior}

Once a spatially-localized Bose clump/soliton emerges, its subsequent evolution and long term dynamics depends on whether the short-ranged self-interactions are attractive or repulsive. The full spectrum of such solitons is extensively discussed in the literature. See e.g.~\cite{Chavanis:2011zi,Chavanis:2011zm,Chavanis:2022fvh,Schiappacasse:2017ham}. To recapitulate some of the basic points that may suffice for our purposes, consider the energy landscape for objects of radius $r_s$ and mass $M_s$ in the theory. Using Eq.~\eqref{eq:Hamiltonian_density}, the wave pressure energy, self-gravitational potential energy, and short-ranged self-interaction potential energy are $H_{\rm wp} = aM_s/(m^2r_s^2)$, $H_{\rm gr} = -b (4\pi G)M_s^2/(r_s)$, and $H_{\rm self} = -c\lambda M_s^2/(m^4 r_s^3)$ respectively, with $a$, $b$, and $c$ some positive coefficients that depend upon the exact profile of the object. The total energy is the sum of all three.

\subsubsection{Attractive short-ranged interactions $\lambda > 0$}

For this case, the energy vs radius curve (for a given mass $M_s$) has a \textit{local} minima that corresponds to quasi stable negative energy (bound) states / solitons. It is separated from the runaway behavior towards small radii, $\sim -\lambda/r_s^{3}$, by a barrier whose height decreases with increasing $M_s$. The barrier disappears at a critical mass $M_{s,\rm crit} \propto (\lambda G)^{-1/2}$, beyond which the theory does not admit any quasi-stable bound states anymore. Starting in the kinetic regime and upon relaxation, a quasi-stable Bose clump nucleates and starts to accrete mass from its surroundings. Ultimately once it accumulates enough mass such that the energy barrier gets sufficiently low, and/or it `breathes' rapidly enough so as to be able to probe beyond the energy barrier, it `collapses'. This is because the region now prefers to lower its energy by transitioning on the runaway $\sim r_s^{-3}$ curve. This is sometimes referred to as ``Bosenova" (owing to its analogy with a type II supernova). While subsequent evolution of the object beyond this criticality requires a fully relativistic analysis and has been pursued in the literature~\cite{Levkov:2016rkk} (also see~\cite{Fox:2023aat} for an associated astrophysical phenomenology), the evolution leading upto this criticality (and even beyond until the wave pressure starts to become comparable to rest mass energy) is well captured by the non-relativistic treatment. For large attractive strengths $\lambda \gtrsim -\lambda_{\rm cr}$, the barrier is less significant and the objects quickly collapses upon nucleation. In our simulations we indeed observe this phenomenon (see fig.~\ref{fig:snapshots} in sec.~\ref{sec:simulations} ahead).

\subsubsection{Repulsive short-ranged interactions $\lambda < 0$}

In the repulsive scenario there is no runaway domain since the energy for low radii is now $-\lambda/r_s^3 > 0$. This renders the previous local minima stable (hence now \textit{global} minima), corresponding to bound soliton states. The critical mass $M_{s,\rm crit} \propto (-\lambda G)^{-1/2}$ serves as the transition point into the Thomas-Fermi regime~\cite{Boehmer:2007um,Chavanis:2011zi,Magana:2012ph}. This is where the mass of solitons gets large enough to admit comparable amounts of self-gravitational and short-ranged interaction energy densities, with gradient pressure becoming sub-dominant. As a result, the radius starts to approach a constant $r_{\rm s} \sim \sqrt{-\lambda}/(m^2\sqrt{4\pi G})$ (with the mass dependent correction term dying out as $\sim M_s^{-1}$). Up until the mass becomes sufficiently large where $GM_{s,\rm relv} \sim r_s$ and relativistic effects start to become important (see~\cite{Salehian:2021khb,Croon:2018ybs}), the theory then admits a set of ``Chandrasekhar solitons" with masses ranging anywhere between $M_{s,\rm crit}$ and $M_{s,\rm relv}$, and radii approximately around $r_{\rm s} \sim \sqrt{-\lambda}/(m^2\sqrt{4\pi G})$.\footnote{The reason we call them ``Chandrasekhar" solitons (also see~\cite{Jain:2022kwq}) is because of the scaling of their maximum mass $M_{s,\rm relv}$. It behaves similarly to that of the Chandrasekhar limit for degenerate stars $M \propto G^{-3/2}m^{-2}$, and can be attributed to the fact that the Fermi pressure essentially gets replaced with repulsive short-ranged self-interactions~\cite{Boehmer:2007um,Chavanis:2011zi,Magana:2012ph,Hertzberg:2020xdn}.}

Though the theory admits these stable Chandrasekhar solitons, understanding their evolution and long term behavior within the bath of DM waves is crucial, and has been extensively studied in the literature. See~\cite{Dawoodbhoy:2021beb,Shapiro:2021hjp,Hartman:2021upg,Hartman:2022cfh,Foidl:2023vat} for simulation setups using the fluid/Madelung equations instead of the \schr field equation for scalar wave dark matter (with repulsive short-ranged self-interaction). For our Fourier split simulation technique (which we discuss in the next section), we find that over \textit{longer} time scales (after the nucleation of Bose clumps), the system reaches some sort of criticality when high frequency modes (near cutoff) start to appear in the simulation box. This leads to a breakdown of the simulation (along with the disruption of the clump), visible in the form of a checkerboard-like pattern. We present this peculiar artifact from our simulations in appendix~\ref{sec:peculiarity}, although a detailed investigation of it is left for future work.

\section{Field simulations}\label{sec:simulations}

To verify our analytical understanding of kinetic relaxation and associated nucleation of bound Bose stars, we have carried out a large suite ($\sim 500$) of 3D simulations of the GP system~\eqref{eq:SP_physical} with varying values of the nGSI strength $\lambda$. We evolve the GP system~\eqref{eq:SP_physical} with 
the following initial Gaussian function for the ${\bm k}$-space \schr field 
\begin{align}\label{eq:Gaussian_initialansatz}
    V^{-1/2}\,\Psi_{\bm k/m}\Bigl|_{t=0} &= e^{i\theta_{\bm k/m}}\sqrt{f_{\bm v}}\Bigl|_{t=0}\nonumber\\
    &= e^{i\theta_{\bm k/m}}\left[\frac{(2\pi)^{3/2}\bar{\rho}}{m(m v_0)^3}\,e^{-\frac{v^2}{2v_0^2}}\right]^{1/2},
\end{align}
where $\theta_{\bm k/m}$ are random phases, uniformly distributed in $(0,2\pi)$, for every wavenumber ${\bm k}$. Our numerical algorithm is based on the well known split Fourier technique/pseudo-spectral method~\cite{Springel:2005mi,Mocz:2017wlg,Edwards:2018ccc,Glennon:2020dxs,Jain:2022agt,Jain:2023qty},
and we have used both Python based and Matlab based codes to generate our simulation data.

As mentioned earlier, in order to be in the kinetic regime we require (i) interactions to be tiny as compared with the typical free wave evolution (occurring over time scales $\sim 2/mv_0^2$), and (ii) the box size to be larger than the typical field fluctuation scale $\pi(mv_0)^{-1}$. Furthermore, we also impose the box size to be smaller than the gravitational Jeans scale associated with a incoherent bound halo $\ell_{J} \sim v_0(\pi/G\bar{\rho})^{1/2}$, in order to avoid its formation within our simulation box. In this sense, our simulation box of a collection of DM waves with typical fluctuation scale $\sim \pi(mv_0)^{-1}$, may be regarded as a region within a DM halo. In summary we require the following to hold true
\begin{align}
    {\rm Kinetic\,regime:} \quad L &\gg \pi(mv_0)^{-1}\quad {\&} \quad \Gamma_{\rm relax} \ll mv_0^2/2\,,\nonumber\\
    {\rm sub\,Jeans\,scale:} \quad L &< \ell_{J} \sim v_0(\pi/G\bar{\rho})^{1/2}\,.
\end{align}
In our simulations, we work with dimensionless quantities, for which purpose we set $G = 1/(8\pi)$ and $m=1$. More explicitly, one can rescale different quantities in the fashion $t \rightarrow t/\mathcal{E}$, ${\bm x} \rightarrow {\bm x}/\sqrt{m\,\mathcal{E}}$, $\psi \rightarrow \psi\,\mathcal{E}/\sqrt{8\pi Gm}$ and $\lambda \rightarrow \lambda\mathcal{E}/(8\pi G m^3)$ to get Eq.~\eqref{eq:SP_physical} with both $8\pi G$ and $m$ replaced by unity.
Here $\mathcal{E}$ is any reference energy scale in the system (for instance $\mathcal{E} = mv_0^2/2$). The discretization in space is simply $\Delta x = L/(N_x-1)$ where $L$ and $N_x^3$ are the box size and number of grid points respectively, and the time discretization is $\Delta t = 2\pi(\Delta x)^2 m/(3\eta)$ with $\eta \geq 1$.\footnote{The $\eta \geq 1$ makes sure that there is at-least one time point in between the full $2\pi$ rotation of the fastest oscillating mode $k_{\rm max} \sim 2\pi/\Delta x$. Since any faithful dynamics of the system should not be sensitive to high frequencies (corresponding to the box discretization scale), $\eta$ can even be smaller than unity. For all our simulations, $\eta$ is at-least as big as unity.} In the split Fourier technique, the field evolution is split into a drift part where it is evolved solely due to the gradient term (free field evolution), and a kick part where it is evolved solely due to interactions.
The Courant-Friedrichs-Lewy (CFL) condition ensures that the fastest process in the dynamics is captured appropriately. Hence, the fastest amongst the kick and drift processes, at any time iteration, sets the time discretization $\Delta t$ (e.g. see~\cite{Jain:2022agt, Jain:2023qty} for details). In the kinetic regime, the time discretization is always set by the free field evolution term $\sim 
(\Delta x)^{-2}/2m$, and hence by space discretization as given above. 

In all of our simulations we set $v_0 = 1/\sqrt{2}$, and choose the box size and average mass density such that we are deep in the kinetic regime. Most of the simulations were performed with $L = 40$, $L = 45$, and $L = 50$ box sizes, and the average mass densities were chosen to be small enough such that the factor $\Gamma^{-1}_{\rm relax}\,mv_0^2/2$ was at-least as large as $\sim 250$, going all the way up-to even $\sim 4500$. For robustness, we have performed simulations with different grid sizes $N_x = \{192,216,256\}$, scanning over different $\lambda/\lambda_{\rm cr}$ values. (We also performed simulations with $N_x = 150$ and $N_x = 300$ to test convergence of our results (see Appendix~\ref{sec:convergence}).\\

To capture the formation of localized Bose clumps, we keep track of the mass density in the box, radially averaged (in $\bm k$ space) occupation number function $f_{k}$, the associated volume averaged correlation function $\zeta(r)$, and the maximum mass density in the box $\rho_{\rm max}$.\footnote{In all of our simulations, we confirm the behavior of $f_k$, in that it develops increasing support towards smaller $k$ values, at-least up until nucleation.} Nucleation of a localized clump can be characterized by a change of trend of $\rho_{\rm max}$, wherein it starts to monotonically increase beyond just the statistical fluctuations that happen over short time scales. We record the corresponding times in all of our simulations, both by direct inspection and statistical methods such as moving average.\footnote{We note that this is not the only way to know whether a bound clump has formed or not. For instance one can alternatively construct an energy spectral function as in~\cite{Levkov:2018kau}, to extract the time scale when the function develops a support towards negative $\omega$.}\\

In order to gauge the validity of our analytical estimate of nucleation times (c.f. Eq~\eqref{eq:taunuc} with Eq.~\eqref{eq:rate_master}), and to extract the different $\alpha$ coefficients, we split the data set into two, with $\lambda = -2\pi G m^2/v_0^2$ being the splitting point. Below we elaborate on the statistical analysis we performed in the two regimes.

\begin{figure*}[!t]
    \centering
\includegraphics[width=0.985\textwidth]{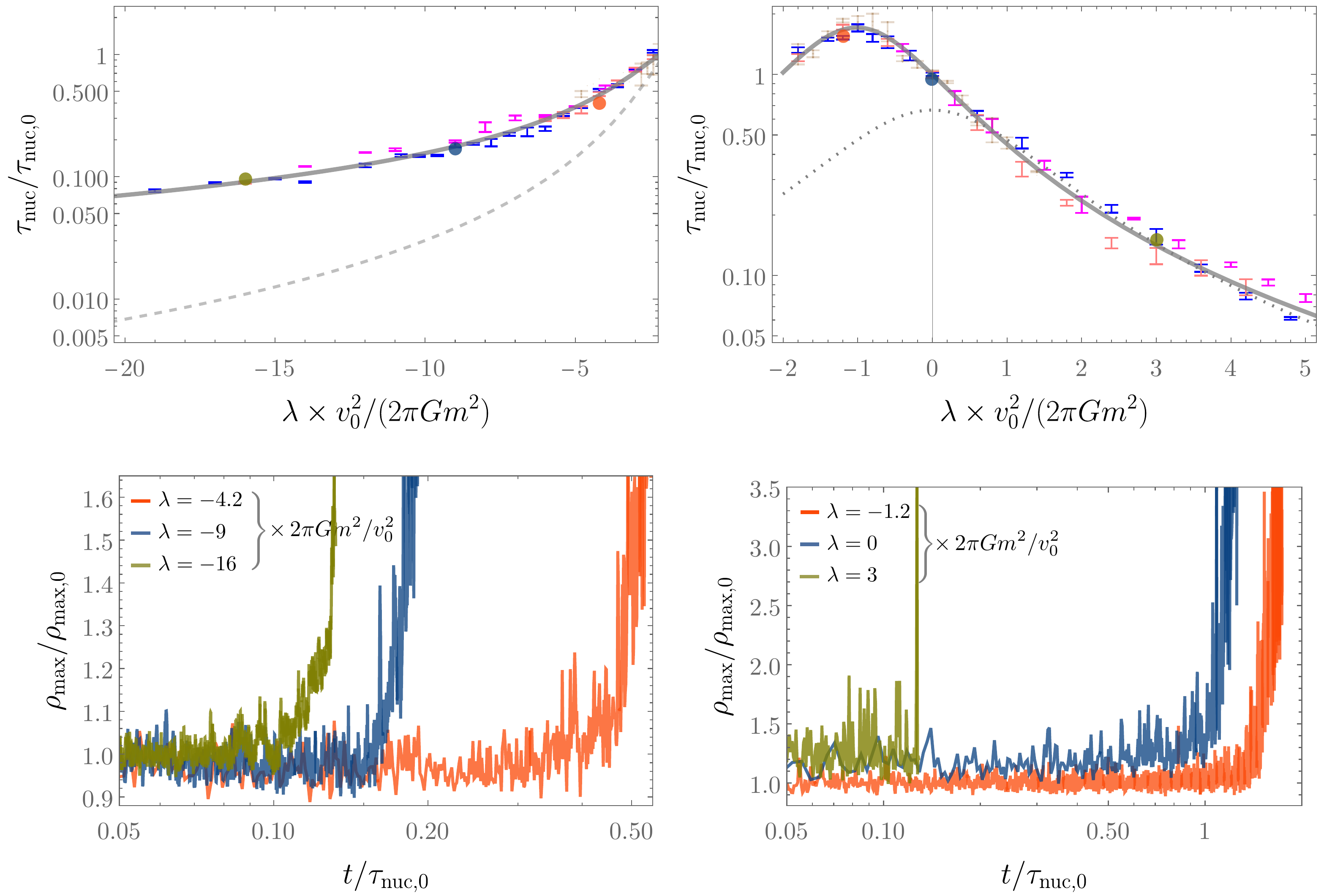}
    \caption{\textit{Upper Panel}: Our main figure showing the nucleation time $\tau_{\rm nuc}$ (normalized by the gravity only case) as a function of short-ranged self-interaction strength $\lambda$ (normalized by the critical factor $2\pi G m^2/v_0^2$). Solid gray curves are from the theory estimate Eq.~\eqref{eq:taunucovertaugr} (c.f. Eq.~\eqref{eq:taunuc} with Eq.~\eqref{eq:rate_master} and Eq.~\eqref{eq:hfunc}), where the different $\alpha$ coefficients are obtained from least square fitting as described in the main text. The different colored $1\sigma$ bars are from simulations (performed with Gaussian initial conditions~\eqref{eq:Gaussian_initialansatz}), with box sizes $L = 36$ (brown), $L = 40$ (pink), $L = 45$ (magenta), and $L = 50$ (blue), and varying average densities $\bar{\rho}$. Here we have only plotted one theory curve for $L = 50$ (all curves for the four different box sizes lie practically on top of each other since the $L$ dependence is quite mild). To show the effect of the interference term in the relaxation rate and the delay factor in the nucleation time, we have also plotted dotted and dashed gray curves. Respectively, these correspond to when the interference term from the relaxation rate is set to zero, and the delay factor in the nucleation time scale is set to unity. This delay factor is only relevant in the $\lambda \lesssim 2\pi G m^2/v_0^2$ case, and the dashed gray curve in the left panel is simply the extension of the main solid gray curve in the right panel.  \textit{Bottom panel}: Normalized $\rho_{\rm max}$ (by their respective initial values) vs time curves for six different $\lambda$ values, highlighted by colored points in the upper panel. The points of `sudden' rise correspond to nucleation of respective localized Bose clumps.} 
    \label{fig:taunuc_n_rhomax}
\end{figure*} 

\subsubsection{Net attractive interactions ($\lambda \gtrsim \lambda_{\rm cr}$).}

In order to test the $\lambda$ dependence of our estimate, we construct the quantity $r(\lambda) = \log(mv_0L)(\tau_{\rm nuc}(0) - \tau_{\rm nuc}(\lambda))/2\tau_{\rm nuc}(\lambda)$ using Eq.~\eqref{eq:rate_master} and Eq.~\eqref{eq:taunuc}. This gets rid of the $\bar{\rho}$ and $L$ dependence, giving
\begin{align}
    r(\lambda) = \frac{\alpha_{12}}{\alpha_{1}}\left(\frac{\lambda\,v_0^2}{2\pi G m^2}\right) + \frac{\alpha_{2}}{2\alpha_{1}}\left(\frac{\lambda\,v_0^2}{2\pi G m^2}\right)^2\,\nonumber\,.
\end{align}
Not only is the curve simple enough to do statistics with, this way we can also combine all of our simulation data (with different $\bar{\rho}$ and $L$). The analogous quantity for simulations is
\begin{align}
    \hat{r}(\lambda) = \Biggl[\frac{\langle\hat{\tau}_{\rm nuc}(0)\rangle - \hat{\tau}_{\rm nuc}(\lambda)}{2\langle\hat{\tau}_{\rm nuc}(\lambda)\rangle}\,\Biggr]\log(m v_0 L)\nonumber\,,
\end{align}
where hats denote simulation data and angle brackets denote averaging over all of the data (for a given $\lambda$ value). To extract the ratios $\alpha_{12}/\alpha_{1}$ and $\alpha_{2}/\alpha_{1}$ for the theory curve, we construct the cost function
\begin{align}\label{eq:costfunction1}
    {\rm cost}\left(\frac{\alpha_{12}}{\alpha_1},\frac{\alpha_{2}}{\alpha_1}\right) = \sum^{\sim \lambda_{\rm cr}}_{\lambda}\frac{1}{N_{\lambda}}\sum^{N_{\lambda}}_{i=1}\Biggl[\frac{\hat{r}_i(\lambda) - r(\lambda)}{r(\lambda)}\Biggr]^2
\end{align}
for least square fitting. Here $N_{\lambda}$ is the number of different simulations performed for a given $\lambda$ value. Minimizing this cost function then fetches the optimal values for $\alpha_{12}/\alpha_{1}$, and $\alpha_{2}/\alpha_{1}$. For $\alpha_1$, we simply find the average of $\hat{\tau}_{\rm nuc}(0)/\tau_{\rm nuc}(0)$, which we then use to get $\alpha_{12}$ and $\alpha_{2}$ from the previous two ratios. For our Gaussian initial condition~\eqref{eq:Gaussian_initialansatz}, we found
$\alpha_{1} \simeq 0.8$,  $\alpha_{12} \simeq 1.2$, and $\alpha_{2} \simeq 1.2$. 

\subsubsection{Net repulsive interactions ($-\lambda \gtrsim -\lambda_{\rm cr}$)}

In this case, we expect nucleation to happen later than relaxation, given by Eq.~\eqref{eq:taunuc} with the delay factor $h(\lambda)$ in Eq.~\eqref{eq:hfunc}. We can use the previous case relationship $\tau_{\rm nuc}(\lambda \gtrsim \lambda_{\rm cr}) \simeq \Gamma^{-1}_{\rm relax}$, to test the scaling of $h(\lambda)$ for the $-\lambda \gtrsim -\lambda_{\rm cr}$ case. From simulations, we construct $\hat{\tau}_{\rm nuc}\Gamma_{\rm relax}$ with the three $\alpha$s in the relaxation rate set to the ones obtained above. We then perform least square fitting by constructing the cost function similar to the previous case
\begin{align}\label{eq:costfunction2}
    {\rm cost}(\alpha_3) = \sum^{\sim \lambda_{\rm cr}}_{\lambda}\frac{1}{N_{\lambda}}\sum^{N_{\lambda}}_{i=1}\Biggl[\frac{\hat{\tau}_{\rm nuc}\Gamma_{\rm relax} - h(\lambda)}{h(\lambda)}\Biggr]^2\,,
\end{align}
and minimizing it. For Gaussian initial conditions, we found $\alpha_{3} \simeq 1$.\\

With the above analysis and all the four $\alpha$ values obtained, upper panel of fig.~\ref{fig:taunuc_n_rhomax} shows our main plot. We plot $\tau_{\rm nuc}(\lambda)$ (normalized by $\tau_{\rm nuc}(0)$) as a function of $\lambda(v_0^2/2\pi G m^2)$:
\begin{align}\label{eq:taunucovertaugr}
    \frac{\tau_{\rm nuc}}{\tau_{\rm nuc,0}}(x) = \frac{2\alpha_1\Lambda\,h(x)}{2\alpha_1\Lambda + 4\alpha_{12}x + \alpha_2x^2}\,;\quad x \equiv \frac{\lambda v_0^2}{2\pi G m^2}\,,
\end{align}
along with our simulation data. Here
$\Lambda = \log(mv_0L)$ is the Coulomb logarithm, and $h(x)$ is unity for $x \gtrsim -1$ (right upper panel of fig~\ref{fig:taunuc_n_rhomax}) while linearly increasing as $-x \gtrsim -1$ (left upper panel of fig~\ref{fig:taunuc_n_rhomax}). Note that we have only plotted one curve for $L = 50$ (solid gray), since the dependence on $L$ is very mild and renders different curves for different values of $L$ practically on top of each other. The error bars correspond to $1$-$\sigma$ fluctuations (owing to random and different initial condition for every simulation seed), with different colors corresponding to three different box sizes considered. In general, the agreement between analytical estimates and simulations is evident. Let us highlight our two main results: (a) The rising feature as $\lambda$ goes from positive to negative, with a peak occurring around $\lambda \simeq -2\pi Gm^2/v_0^2 \simeq \lambda_{\rm cr}$, is a clear evidence of the interference term in the relaxation rate. To represent the effect of the interference term visually, we have also plotted a dotted gray curve (in the upper right panel of fig.~\ref{fig:taunuc_n_rhomax}), which is equal to inverse of the relaxation rate with the interference term dropped. That is, inverse of Eq.~\eqref{eq:rate_master} with the term $\propto G\lambda$ set to zero; (b) To the left of the peak and increasing $-\lambda$, nucleation happens later than just the inverse relaxation rate. The delay factor $h$ and the relaxation rate $\Gamma_{\rm relax}$ scale as $\sim -\lambda$ and $\lambda^2$ (to leading order) respectively, resulting in the scaling of the nucleation time as $(-\lambda)^{-1}$ (and not $\lambda^{-2}$) to leading order. To highlight this, we have augmented the upper left panel of fig.~\ref{fig:taunuc_n_rhomax} with just the relaxation time curve, i.e. $\Gamma_{\rm rel}^{-1}$, shown in dashed gray. (This is nothing but the solid gray curve on the right upper panel, extended towards the left upper panel).

\begin{figure*}[!t]
    \centering
\includegraphics[width=0.975\textwidth]{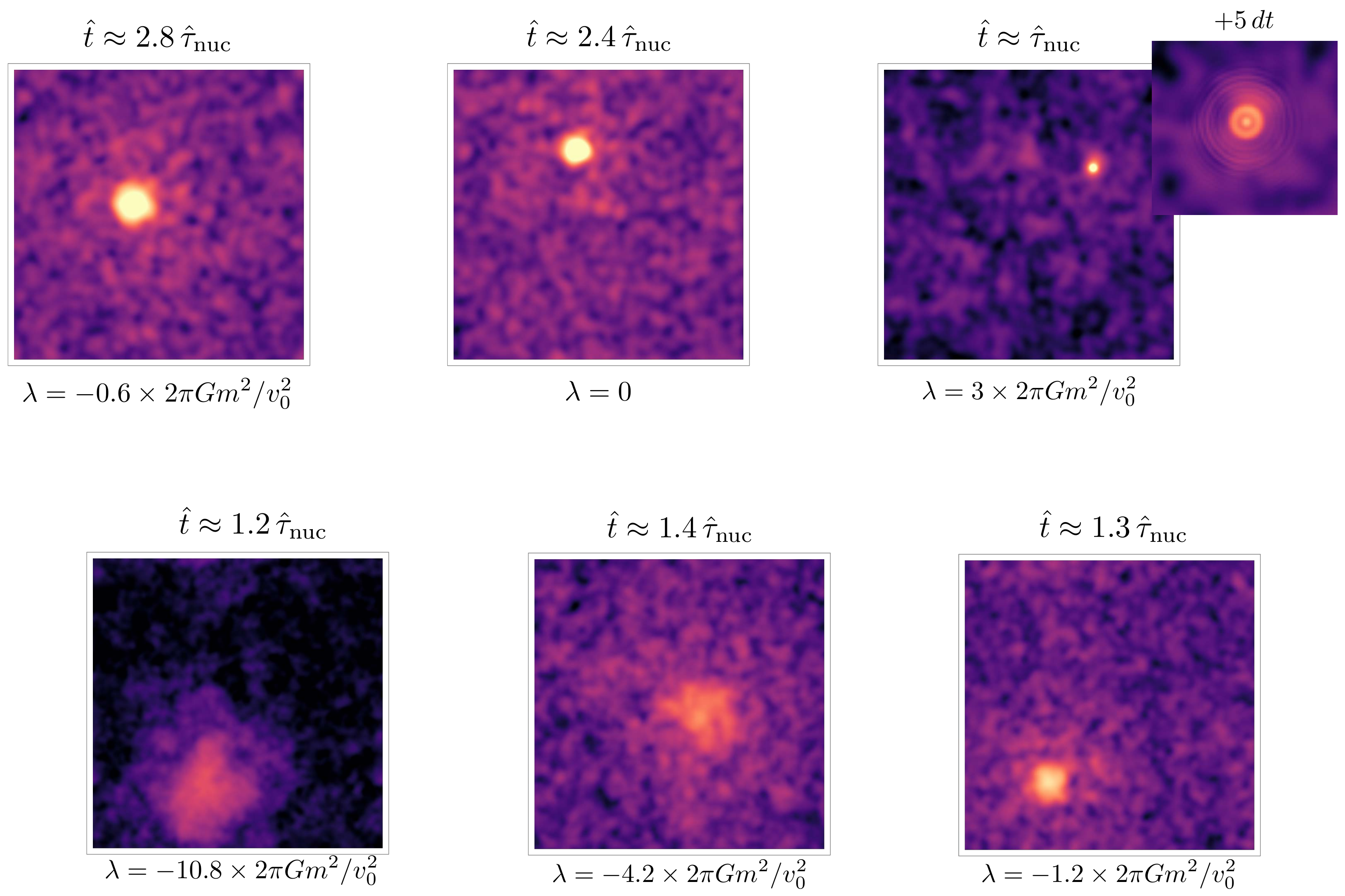}
    \caption{Density projection snapshots for 6 different $\lambda$ values, at different times $\hat{t}$ in the respective simulations. 
    \textit{Upper Panel}: Snapshots for three $\lambda$ values in the typical \textit{net} attractive regime $\lambda \gtrsim \lambda_{\rm cr} \simeq -2\pi Gm^2/v_0^2$. In the rightmost snapshot, for $\lambda \approx 3\,\lambda_{\rm cr}$, the nucleated Bose clump quickly collapses (within $5 dt$), shown in the smaller right corner image.  \textit{Bottom panel}: Snapshots for three different $\lambda$ values for the other case of typical \textit{net} repulsive self-interactions $-\lambda \gtrsim -\lambda_{\rm cr} \simeq 2\pi Gm^2/v_0^2$.} 
\label{fig:snapshots}
\end{figure*} 

The lower panel of fig.~\ref{fig:taunuc_n_rhomax} shows moving averaged $\rho_{\rm max}$ vs time curves for six simulations with different $\lambda$ values. The unambiguous ``sudden" rise in $\rho_{\rm max}$ marks the nucleation of a localized object within which density grows over time.\footnote{In all of our simulations, we have explicitly verified, by visually tracking the simulation box, that this rising feature indeed corresponds to appearance of an overdense region.}


As visual examples, in fig.~\ref{fig:snapshots} we also present density projection snapshots for six different $\lambda$ values at later times, showing the presence of nucleated Bose stars. 
For attractive short-ranged self-interaction $\lambda > 0$, nucleated Bose clumps eventually collapse into a Bosenova. This happens when it reaches the critical mass where it can no longer remain stable (see section~\ref{sec:eventualbehavior}).


\section{Summary and Discussion}\label{sec:summary}

In this paper we have investigated kinetic relaxation and associated nucleation times of Bose stars, in scalar fuzzy dark matter with short-ranged $2$-body self-interactions. Starting with the wave-kinetic Boltzmann equation for the mode occupation number function (which we derived in an earlier work), we first highlighted the presence of a cross/interference term $\propto G\lambda$ in the rate of relaxation $\Gamma_{\rm relax}$, alongside the usual two terms $\propto G^2$ and $\lambda^2$ due to both gravitational and short-ranged self-interaction individually. This is because of the wave-mechanical nature of the system: The rate depends on the total cross section, which is not just the sum of individual cross sections due to the different processes. Rather the scattering amplitudes due to all the processes must be added first, and then use its absolute square to get the cross section and associated rate of relaxation/condensation.

The presence of this cross term gives rise to a critical repulsive self-interaction strength
$\lambda_{\rm cr} \approx -2\pi Gm^2/v_0^2$, around which the typical net self-interaction (due to both gravitational and short-ranged self-interaction), transitions between being attractive and repulsive, and the relaxation rate becomes smallest. Here $k_0 = mv_0$ is the typical wave-mode present in the system initially. 

For nucleation times as a function of $\lambda$, we found that for the case of net attractive self-interaction $\lambda \gtrsim \lambda_{\rm cr}$, nucleation happens quickly upon relaxation, giving rise to the relationship $\tau_{\rm nuc} \simeq \Gamma^{-1}_{\rm relax}$. One the other hand for net repulsive self-interaction $-\lambda \gtrsim -\lambda_{\rm cr}$, nucleation is delayed. This is because upon relaxation, short-ranged self-interaction dominate over gravitational self-interaction, preventing nucleation of a bound object. Over time as more particles are driven towards the condensate phase (equivalently, as field correlation length scale increases along with diminishing density fluctuations), a potential arises for the formation of a bound region where gravity can now overcome both the wave pressure and short-ranged self-interaction. The associated delay factor rises linearly with $-\lambda$, giving the nucleation time scale as $\tau_{\rm nuc} \simeq (\lambda/\lambda_{\rm cr})\Gamma^{-1}_{\rm relax}$. In summary, Eq.~\eqref{eq:taunuc} along with Eq.~\eqref{eq:rate_master} (with the delay factor give in Eq.~\eqref{eq:hfunc}) is our main analytical estimate for the nucleation timescale of Bose stars, as a function of the short-ranged self-interaction strength $\lambda$.\\

To analyze this, we performed a large suit of $3$+$1$ dimensional simulations of the \schr-Poisson / Gross-Pitaevskii system (Eq.~\eqref{eq:SP_physical}), for many different values of $\lambda$ and different parameters such as the box size $L$ and average mass density $\bar{\rho}$. All of our simulations were carried out with Maxwell-Boltzmann distribution, with random phases for each value of the wavemode ${\bm k}$ (Eq.~\eqref{eq:Gaussian_initialansatz}). Throughout most of our simulations, we kept track of the max density in the box, occupation number function $f_{k}$ (radially averaged $f_{\bm k}$ in ${\bm k}$ space), the associated correlation function $\zeta (r)$, and projected mass density along some direction. By reading the times at which $\rho_{\rm max}$ starts to monotonically rise beyond just the statistical fluctuations (together with making sure that a localized over dense region does appear in the simulation box around this time), we record the times of nucleation. Upper panel of Fig.~\ref{fig:taunuc_n_rhomax} presents the comparison between simulations and analytical estimate. As examples, the figure is also appended (lower panel) with $\rho_{\rm max}$ vs time curves for six different $\lambda$ values.\\

While in this paper we have not analyzed our simulation data for the rate at which Bose stars accrete mass from their surroundings, we kept track of the eventual behavior of these objects (post nucleation), for many of our simulations. For the attractive case ($\lambda > 0$), we confirmed that the nucleated Bose stars eventually decay away. This is expected since there exists a maximum critical mass beyond which the star becomes unstable and collapses into a Bosenova. For instance see upper right snapshot in Fig.~\ref{fig:snapshots}, when the nucleated star `immediately' collapses. While the study of eventual dynamics and fate of such regions requires a full relativistic treatment, field dynamics up-to this point is well described by the non-relativistic GP equation (e.g. see~\cite{Levkov:2016rkk}).

For the repulsive case $\lambda < 0$, we found a peculiar decay behavior. We find that the nucleated clump \textit{eventually} (over time scales longer than the nucleation time) reaches a type of criticality at which point very high frequency modes, passing through the clump and travelling along the three directions of the simulation box, appear in the system. See appendix~\ref{sec:peculiarity} for some discussion. This could be an artifact of the periodic boundary conditions of the split-Fourier simulation setup, and if so, bringing into question its use to study long term dynamics of fuzzy dark matter with repulsive short-ranged self-interactions via such simulation setups. We leave a detailed investigation of this behavior for a separate work.

\subsection{Comparison with earlier work}

Let us now compare our results with some of the earlier work on the subject of kinetic nucleation of Bose stars. First, our results encompass the result of~\cite{Levkov:2018kau} for the gravity only ($\lambda = 0$) case, and is even in very good agreement with the order unity coefficient $\alpha_1$ in the rate expression (besides the overall scaling with $\bar{\rho}$, $m$, $v_0$, $L$ and $G$), obtained for Gaussian initial condition. Upon inclusion of short-ranged self-interaction ($\lambda \neq 0$ case), our results differ significantly from the existing literature~\cite{Kirkpatrick:2020fwd,Chen:2021oot,Kirkpatrick:2021wwz}. First, we find that there exists an interference term $\propto G\lambda$ in the \textit{relaxation} rate, which in fact is the leading order $\lambda$ dependent term when short-ranged self-interaction is not dominating over gravitational self-interaction. Only in the scenario when the former is dominant, does the relaxation rate goes as $\lambda^2$ to leading order. Secondly, the \textit{nucleation} time scale is not always equal to the inverse relaxation rate. While for $\lambda \gtrsim \lambda_{\rm cr}$, nucleation time scale is just the inverse relaxation rate, for the strong repulsive self-interaction $-\lambda \gtrsim -\lambda_{\rm cr}$, nucleation time is delayed by an extra factor of $(\lambda/\lambda_{\rm cr})$. Therefore for the purposes of nucleation of Bose stars, only in the case of strong attractive short-ranged self-interaction, $\lambda \gtrsim -\lambda_{\rm cr}$, is it true that the nucleation time goes as $\lambda^{-2}$ to leading order. For in the opposite case of strong repulsive short-ranged self-interaction, $-\lambda \gtrsim -\lambda_{\rm cr}$, the nucleation time scale goes as $\lambda^{-1}$ to leading order instead.


\subsection{Implications}

Our results could have important implications in the context of self-interacting fuzzy dark matter and various interesting phenomenon that it entails. The appearance of the interference term in the relaxation rate, and hence in the nucleation time scale of Boson stars, may modify results for some of the phenomenon  such as recurrent axinovae~\cite{Fox:2023aat}, de-stabilization of gravitational atoms~\cite{Budker:2023sex}, among others. 

In general, irrespective of the nature (attractive or repulsive) of point-like self-interaction, the interference term becomes the leading order $\lambda$ dependent term (and hence extremely important), when $|\lambda|$ is at best comparable to the critical value $|\lambda_{\rm cr}|$. As an example, even for the QCD axion we have 
$\lambda_{\rm qcd}/|\lambda_{\rm cr}| \simeq 1.3(v_0^2 \mpl^2/f_a^2)$, hence becoming comparable to, or less than $|\lambda_{\rm cr}|$, in cosmological environments with $v_0 \lesssim (f_a/\mpl) \sim 10^{-5}$. 
For instance this could be important in the study of axion miniclusters~\cite{Kolb:1993zz}.\\

In this paper we have focused on kinetic nucleation via both gravitational and short-ranged self-interactions for a single scalar field. A natural generalization is to include multiple scalar fields with naturally different masses and 4-point interactions, or a single spin-$1$ field including density-density and spin-spin interactions~\cite{Zhang:2021xxa,Jain:2022kwq}, or even multiple spin-$1$ fields with extra Yang-Mills interactions~\cite{Jain:2022kwq}, or a combination thereof. While there would necessarily be interference terms $\propto G\lambda$, and we expect similar scaling of nucleation time scales as presented in this work (as a function of $\lambda$), a detailed analysis of such cases is left for future work.\footnote{We thank Benjamin Schussler for carrying out some preliminary simulations for the self-interacting vector case, confirming the presence of the interference term in the relaxation rate.}

\section*{Acknowledgements}
We thank Mustafa Amin, Mark Hertzberg, Andrew Long, and David J.E. Marsh for many helpful discussions and also their comments on this manuscript. MJ is partially supported by a DOE grant DE-SC0021619, and partly supported by a Leverhulme Trust Research Project (RPG-2022-145). JT and WW acknowledge undergraduate summer support from the Department of Physics and Astronomy at Rice University.

\bibliography{reference}

\appendix

\section{Statistical convergence}
\label{sec:convergence}

\begin{figure}[!t]
    \centering
\includegraphics[width=0.48\textwidth]{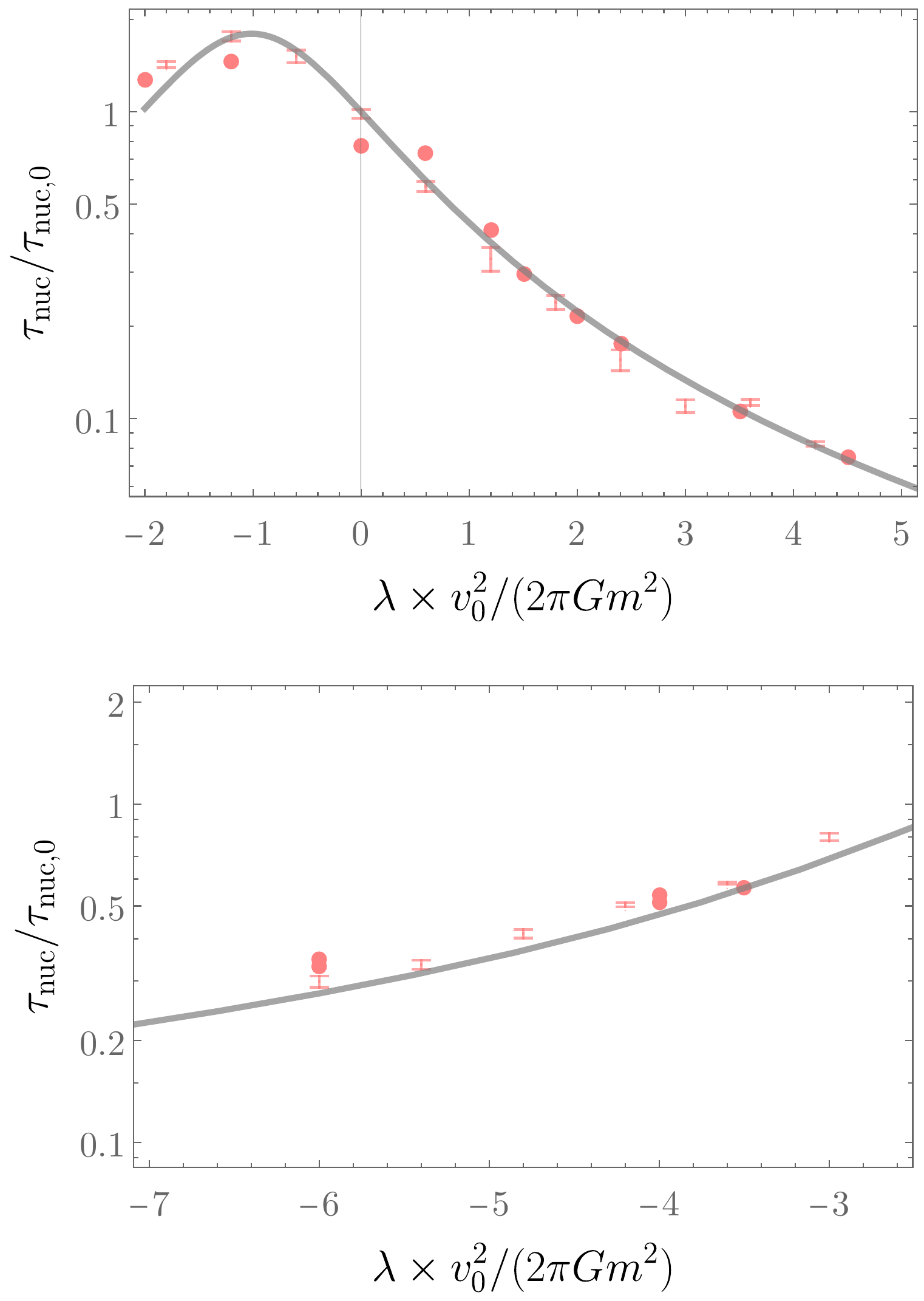}
    \caption{Similar to the upper panel of figure~\ref{fig:taunuc_n_rhomax}, for simulations performed with lower and higher resolution grids compared to the ones used in the main text. With box size $L=40$, the $1\sigma$ bars represent simulations performed using $150^3$ grid, while the solid points are from simulations using $300^3$ grid. The solid gray curves correspond to the analytical estimate Eq.~\eqref{eq:taunucovertaugr} (c.f. Eq.~\eqref{eq:taunuc} with Eq.~\eqref{eq:rate_master} and Eq.~\eqref{eq:hfunc}), with the $\mathcal{O}(1)$ $\alpha$ coefficients obtained using simulations performed with grid sizes $192^3$, $216^3$, and $256^3$ as in the main text.} 
    \label{fig:taunuc_n_rhomax_converge}
\end{figure}

Here we show convergence and reliability of our simulation results. Figure~\ref{fig:taunuc_n_rhomax_converge} compares the main theory curve (in solid gray), with data points for the lower ($N_x = 150$) and higher ($N_x = 300$) resolution grids, compared to $N_x = 192$, $N_x = 216$ and $N_x = 256$ used for our results presented in the main text. See caption for details. Convergence of our results is evident from this plot. Note that this is not the usual convergence, where two or more simulations with similar initial conditions are performed, with different values of $\Delta x$ and $\Delta t$. Rather, here we show a `statistical convergence' of sorts. Also, since $\Delta t \propto (\Delta x)^2$, increasing grid size reduces both $\Delta x$ and $\Delta t$.

\section{Peculiar appearance of high frequency modes for the repulsive case}
\label{sec:peculiarity}

In the case of repulsive short-ranged self interaction $\lambda < 0$, we find a peculiar behavior over \textit{long time scales} (later than nucleation). There starts to appear high frequency modes (near cutoff of the simulation box) that pass through the clump and in all the three perpendicular directions of the box. This leads to disruption of the simulation (manifesting in the form of checkerboard like pattern) and of the clump, as these waves circulate within the periodic simulation box. An artifact of this is wiping out of density fluctuations and eventual homogenization of the simulation box.\\

\begin{figure}[!t]
    \centering
\includegraphics[width=0.47\textwidth]{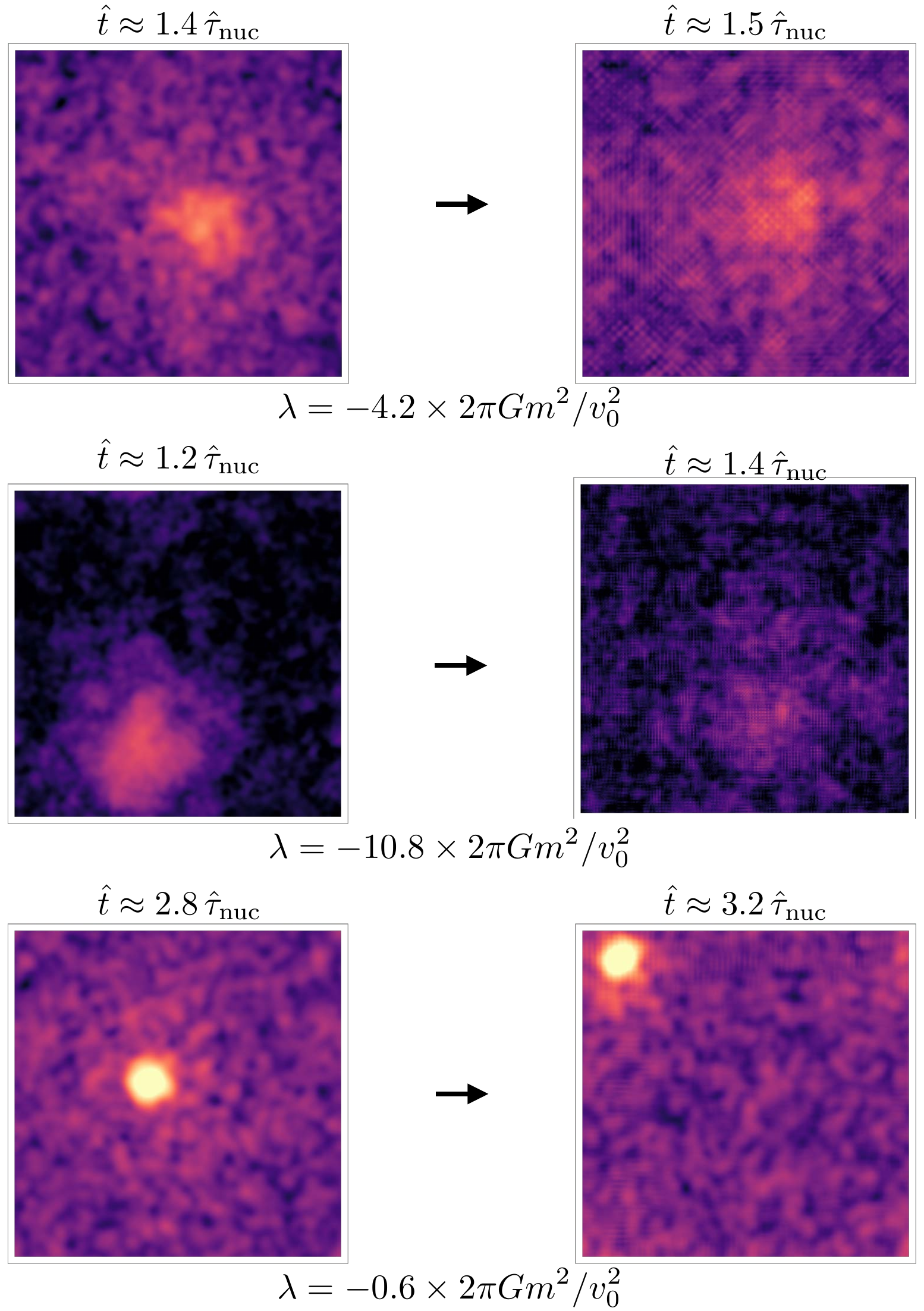}
    \caption{Density projection snapshots for 3 values or repulsive short-ranged self-interaction strength $\lambda$, at two different times in the respective simulations. The grid size is $216^3$, and box length is $50$. The left hand side snapshots are the same as presented in the main text in fig.~\ref{fig:snapshots}. The right hand side snapshots are at later times, when the simulation has now been rendered `unfaithful'. Notice the checkerboard like pattern in all of the snapshots in the right panel.} 
\label{fig:snapshots_peculiar}
\end{figure}

In fig.~\ref{fig:snapshots_peculiar} we provide some simulation snapshots of this peculiarity, for three different values of $\lambda$. Notice the appearance of checkerboard like pattern in the right hand side panel. 

In order to check if the phenomenon is an artifact of finite discretization, we simulated lower and higher resolution grids with the same initial conditions for long times. We found no clues if this is the case or not. As an example, for $\lambda \approx -4.2\lambda_{\rm cr}$, with $L = 40$ and $\bar{\rho} = 0.01$, in the lower resolution ($128^3$) and higher resolution grids ($256^3$) the nucleation times were $\sim 3600$ and $\sim 3800$ respectively. This shows decent convergence of the nucleation time. At the same time however, the onset of these high frequency waves for both the grids were also similar ($\sim 5300$ and $\sim 5700$ respectively).

We also performed two other tests to see if something can be learned about this phenomenon. In one test we put a pre-computed soliton in a bath of DM waves. We observed the same appearance of high frequency modes and checkerboard pattern developing, leading to disruption of the soliton. The times at which this happens depended upon the average mass density and total mass of the soliton. In yet another test, we performed a few runs with larger average mass densities, such that the associated gravitational Jeans scale $\ell_{J} \sim v_0(\pi/G\bar{\rho})^{1/2}$ became smaller than the box size. Even in this case, upon formation of a halo, the same peculiar phenomenon appears. 

While we have confirmed the presence of this feature in our (split-Fourier technique based-)simulations through multiple tests, we leave a detailed analysis of this peculiarity for future work. 

\end{document}